\documentclass[journal,twocolumn,10pt,twoside]{IEEEtran}
\usepackage{graphicx} 

\usepackage{acronym}
\usepackage{tikz}
\usepackage{amsmath,dsfont}
\usepackage{enumitem}
\usepackage{amssymb}
\usepackage[bookmarks,colorlinks]{hyperref}
\usepackage{cite}
\newtheorem{theorem}{Theorem}

\newtheorem{lemma}[theorem]{Lemma} 
\newtheorem{definition}[theorem]{Definition}

\newcommand{\includefig}[1]{\includegraphics[width = \columnwidth]{#1} 	\vspace{-0.2cm}}

\newcommand{\myVec}[1]{{\boldsymbol{#1}}}
\newcommand{\myMat}[1]{{\boldsymbol{#1}}}
\newcommand{\mySet}[1]{\mathcal{#1}}

\newcommand{\E}{\mathbb{E}}		 			
\newcommand{\myX}{{\myVec{x}}}			 	
\newcommand{\myZ}{{\myVec{z}}}			 	
\newcommand{\myS}{{\myVec{s}}}			 	
\newcommand{\LmmseMat}{\myMat{\Gamma}}		
\newcommand{\lenX}{n}			 			
\newcommand{\lenZ}{p}			 			
\newcommand{\lenS}{k}			 			
\newcommand{\Quan}[2]{Q_{#1}^{#2}}			
\newcommand{\myI}{{\myMat{I}}}			 		
\newcommand{\Rate}{R}			 		
\newcommand{\TilM}{\tilde{M}}
\newcommand{\DynRange}{\gamma}
\newcommand{\DynInt}{\Delta}
\newcommand{\myA}{\myMat{A}}
\newcommand{\myB}{\myMat{B}}
\newcommand{\CovMat}[1]{\myMat{\Sigma}_{#1}}
\newcommand{\opt}{^{\rm o}}			
\newcommand{\myEta}{\eta}
\newcommand{\MyKappa}{\kappa}
\newcommand{\eig}[1]{\lambda_{#1}}			
\newcommand{\Wlevel}{\zeta}

\newcommand{\CB}{\mySet{Q}}
\newcommand{\Pdf}[1]{f_{ #1}}

\acrodef{adc}[ADC]{analog-to-digital convertor}
\acrodef{cs}[CS]{compressed sensing}
\acrodef{dtft}[DTFT]{discrete-time Fourier transform}
\acrodef{dft}[DFT]{discrete Fourier transform}
\acrodef{dnn}[DNN]{deep neural network} 
\acrodef{csi}[CSI]{channel state information}
\acrodef{map}[MAP]{maximum a-posteriori probability}
\acrodef{snr}[SNR]{signal-to-noise ratio}
\acrodef{bs}[BS]{base station} 
\acrodef{iot}[IOT]{Interent of Things}
\acrodef{mimo}[MIMO]{multiple-input multiple-output}
\acrodef{mse}[MSE]{mean-squared error}
\acrodef{mmse}[MMSE]{minimum \ac{mse}}
\acrodef{pdf}[PDF]{probability density function}
\acrodef{rv}[RV]{random variable}
\acrodef{fec}[FEC]{forward error correction}
\acrodef{rs}[RS]{Reed-Solomon}
\acrodef{lti}[LTI]{linear time-invariant}
\acrodef{wss}[WSS]{wide-sense stationary}
\acrodef{psd}[PSD]{power spectral density}
\acrodef{ser}[SER]{symbol error rate} 
\acrodef{ber}[BER]{bit error rate} 
\acrodef{sgd}[SGD]{stochastic gradient descent} 
\acrodef{isi}[ISI]{intersymbol interference}  
\acrodef{awgn}[AWGN]{additive white Gaussian noise} 
\acrodef{ut}[UT]{user terminal} 
\acrodef{mmw}[mmWave]{millimeter wave}
\acrodef{cfl}[CFL]{clustered \ac{fl}} 
\acrodef{fl}[FL]{federated learning}
\acrodef{pic}[PIC]{principal inertia compoenent}
\acrodef{ml}[ML]{machine learning} 
\acrodef{dma}[DMA]{dynamic metasurface antenna} 
\acrodef{doa}[DOA]{direction of arrival} 
\acrodef{vqvae}[VQ-VAE]{vector quantization variational autoencoder}

\definecolor{NewColor}{rgb}{0,0,0}

\title{Information Compression in the AI Era: Recent Advances and Future Challenges}
\author{Jun Chen, 
Yong Fang, Ashish Khisti, Ayfer \"Ozg\"ur, Nir Shlezinger, Chao Tian}
\date{April 2024}

\begin{document}

\maketitle

\begin{abstract}
This survey articles focuses on  emerging connections between the fields of machine learning and data compression. While fundamental limits of  classical (lossy) data compression are established using rate-distortion theory, the connections to machine learning have resulted in new theoretical analysis and application areas. We survey recent works on task-based and goal-oriented compression, the rate-distortion-perception theory and compression for estimation and inference. Deep learning based approaches also provide natural data-driven algorithmic approaches to compression. We survey recent works on applying deep learning techniques to task-based or goal-oriented compression, as well as image and video compression. We also discuss the potential use of large language models  for text compression. We finally provide some directions for future research in this promising field.     
\end{abstract}

\section{Introduction}

Data compression has been one of the central technologies in  modern digital world. Two common approaches to data compression are lossy data compression and lossless data compression. In lossy data compression, which is the main focus of this survey, the reconstruction signal need not be  be identical to the source signal, but should satisfy a certain fidelity constraint.  The rate-distortion theory introduced by Shannon~\cite{shannon1948mathematical} establishes fundamental limits of  lossy data compression. While Shannon's framework assumes an idealistic setting where encoding applies over very large blocks of independent and identically distributed (i.i.d.) source samples,  it has nevertheless provided a strong motivation to many practical systems~\cite{berger1998lossy}. Several advances have been made to remove the idealistic assumptions in Shannon's formulation. Finite blocklength analysis of rate-distortion theory~\cite{kostina2016nonasymptotic, wang2011dispersion} provides insights into the gap to Shannon's rate-distortion function when the compression system is operating over finite block-length rather that the asymptotic limit.  More recently the so-called one-shot approach to rate-distortion theory has been studied in a number of works~\cite{elkayam2020one,li2018strong,yassaee2013technique}. These works establish fundamental limits for compressing one source sample at a time. Quite remarkably Shannon's asymptotic analysis can be a good approximation even in such settings, again validating the practical importance of rate-distortion theory.

Advances in artificial intelligence and deep learning have revolutionized many scientific disciplines in the past decade. Generative models e.g., variational auto-encoders, generative adversarial networks and diffusion models, can generate new samples of high quality images that match the distribution of the training dataset. Incorporating these models into lossy data compression enables us to generate realistic looking reconstructions, especially at very low bit-rates where traditional compression codecs produce a number of artifacts~\cite{testolina2021performance}. This has resulted in an interesting interplay between information-theoretic analysis and deep learning based compression techniques. On one hand, the information theoretic analysis has to be revisited to account for {\em realism} constraints in the reconstruction, leading to the study of the rate-distortion-perception tradeoff~\cite{blau2019rethinking}, which is an natural extension of Shannon's rate-distortion tradeoff. On the other hand information theoretic analysis can also yield several architectural insights into neural based compression systems that can simplify the overall implementation~\cite{agustsson2023multi}. Rate-distortion-perception theory  is just one example where deep learning based architectures successfully model semantic information into compression. The emerging field of task-based compression considers design of compression systems where the ultimate objective is to perform efficient compression aimed towards  one or more downstream tasks. As one application, in autonomous driving, capturing a realistic range of rare events with current methods requires around 3 trillion GB of data. At these scales, data are only processed by task-specific algorithms, and storing data in human-readable formats can be prohibitive. Thus extreme levels of compression that are tailored towards downstream tasks (e.g., classification, prediction) need to be developed~\cite{dubois2021lossy}. 

Compression techniques are also crucial in distributed and federated learning~\cite{gafni2021federated}. In such systems multiple client nodes provide training data that must reside on their individual device and not be directly shared with a central entity~\cite{kairouz2021advances}. These nodes participate in collaborative training with a central node. In each round, the central node broadcasts model parameters to all client nodes, which then run model updates based on their local datasets. The central node in turn updates the global model for the next iteration. The transmission of model updates between the individual client nodes and the central nodes can consume a significant amount of bandwidth. The use of compression techniques to reduce bandwidth for model updates has become an active area of research. Compression techniques also play an important role  in both  training and inference involving large foundation models which much be distributed across multiple compute nodes due to memory bottlenecks~\cite{bajic2021collaborative}.

This article provides a survey of these recent advances in information compression, arising from the growing developments in artificial intelligence and machine learning tools.  
In Section~\ref{sec:Comp}, we discuss the theoretical analysis of compression techniques in emerging topics within the field of machine learning. Goal-oriented compression techniques are examined in Section~\ref{ssec:GOQaunt}, while Section~\ref{subsec:rdp} presents recent advances in rate-distortion-perception theory. In Section~\ref{subsec:est}, we explore the role of compression in distributed estimation and learning problems, including its applications in federated learning. The application of machine learning techniques to enhance data compression is surveyed in Section~\ref{sec:AI-Enhanced}. Specifically, we discuss deep learning-based goal-oriented compression in Section~\ref{subsec:Neural-Goal}, the use of large language models (LLMs) for lossless data compression in Section~\ref{subsec:llm}, and the role of deep learning in image and video compression in Section~\ref{sec:video}. We conclude with a summary and future directions in Section~\ref{sec:concl}.


\section{Compression Theory Beyond Rate-Distortion}
\label{sec:Comp}

\subsection{Goal-Oriented Quantization}
\label{ssec:GOQaunt}

We begin by briefly reviewing the standard quantization setup, and recall the definition of a quantizer:
\begin{definition}[Quantizer]
	\label{def:Quantizer}
	A quantizer $\Quan{M}{n,k}\left(\cdot \right)$ with $\log M$ bits, input size $n$, input alphabet $\mySet{X}$, output size $k$, and output alphabet $\hat{\mySet{X}}$ consists of: 
	{\em 1)} An  encoding function $g_n^{\rm e}: \mySet{X}^n \mapsto \{1,2,\ldots,M\} \triangleq \mySet{M}$ which maps the input into a discrete index.
	{\em 2)} A decoding function  $g_k^{\rm d}: \mySet{M} \mapsto \hat{\mySet{X}}^k$ which maps each index $j \in \mySet{M}$ into a codeword $\myVec{q}_j \in  \hat{\mySet{X}}^k$. 
\end{definition}

\begin{figure}
	\centering
	\includefig{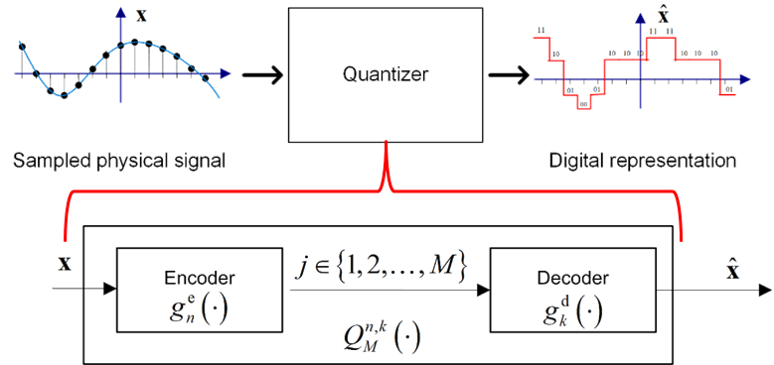} 
	\caption{Quantizer illustration.} 
	\label{fig:Quantizer1}
\end{figure}

We write the output of the quantizer with input $\myX  \in \mySet{X}^n$ as 
\begin{equation*}
 \hat{\myX}  = g_k^{\rm d}\left( g_n^{\rm e}\left( \myX \right) \right) \triangleq \Quan{M}{n,k}\left( \myX \right).   
\end{equation*}
{\em Scalar quantizers} operate on a scalar input, i.e., $n=1$ and $\mySet{X}$ is a scalar space, while {\em vector quantizers} have a multivariate input.  An illustration of a quantization system is depicted in Fig. \ref{fig:Quantizer1}.

In the standard quantization problem, a $\Quan{M}{n,n}\left(\cdot \right)$ quantizer is designed to minimize some distortion measure  $d:\mySet{X}^n\times\hat{\mySet{X}}^n \mapsto \mySet{R}^+$  between its input and its output. 
The performance of a quantizer is  characterized using its quantization rate $\Rate \triangleq \frac{1}{n}\log M$, and the expected distortion $\E\{d\left(\myX , \hat{\myX}  \right)\}$. For a fixed input size $n$ and codebook size $M$, the optimal quantizer is
\begin{equation}
\label{eqn:OptQuantizer}
\Quan{M}{n, {\rm opt}}\left(\cdot \right) = \mathop{\arg \min}_{\Quan{M}{n,n}} \E \left\{d\left(\myX, \Quan{M}{n,n}\left( {\myX} \right)\right)   \right\}.
\end{equation}
Characterizing the optimal quantizer and its trade-off between distortion and quantization rate is in general a very difficult task. Optimal quantizers are thus typically studied assuming either high quantization rate, i.e., $\Rate \rightarrow \infty$, see, e.g., \cite{gray1998quantization}, or asymptotically large inputs, namely, $n \rightarrow \infty$, via rate-distortion theory \cite[Ch. 10]{cover2012elements} (see also Section \ref{subsec:rdp}).

As opposed to standard quantization, in {\em goal-oriented quantization} \cite{zou2022goal}, also referred to {\em task-based quantization}~\cite{shlezinger2018hardware,bernardo2023design},  the design objective of the quantizer is some task (such as minimizing the distortion between a parameter underlying the input and its representation) other than minimizing the distortion between its input and output. In this section, we describe existing approaches to design goal-oriented quantization based on knowledge of the underlying statistical model relating the observations and the task. To that aim, we first discuss the high level rationale of the considered design approaches, after which we formulate the considered setup mathematically. Based on this formulation, we propose dedicated goal-oriented quantization system designs for tasks that can be represented as linear and quadratic functions, respectively. 

\subsubsection{High Level Rationale}
We focus on the generic task of acquiring a random vector  $\myS \in \mySet{S}^\lenS \subseteq \mySet{R}^\lenS$ from a  statistically dependent random vector $\myX \in \mySet{R}^\lenX$ of larger dimensionality, i.e., $\lenX \geq \lenS$.  The set $\mySet{S}$ represents the possible values of the unknown vector.
 The recovered estimate of $\myS$, denoted $\hat{\myS}$, is compressed into a digital representation using up to $\log M$ bits, dictating the bit budget allowed for goal-oriented quantization. The observed $\myX$ is related to $\myS$ via a conditional probability measure $P_{\myX|\myS}$. 

 The performance limits of goal-oriented quantization with asymptotically large vectors, i.e., when $\lenX \rightarrow \infty$ while $\Rate = \frac{1}{n}\log M$ remains fixed, can be characterized using indirect rate-distortion theory \cite{witsenhausen1980indirect}. Specifically, for estimation tasks with the \ac{mse} distortion objective, i.e., $d(\myVec{s}, \hat{\myVec{s}}) = \| \myVec{s} - \hat{\myVec{s}} \|^2 $, the goal-oriented quantization mapping which minimizes the \ac{mse} for a fixed quantization rate $\Rate$ was derived in \cite{wolf1970transmission} for fixed-size vectors. The resulting optimal strategy consists of applying vector quantization to the \ac{mmse} estimate of  $\myVec{s}$ from  $\myVec{x}$. While the optimal system utilizes vector quantization, the fact that such pre-quantization processing can improve the performance in estimation tasks was also demonstrated in \cite{shlezinger2018hardware}, which considered scalar quantizers. However, it was also shown in \cite{shlezinger2018hardware,salamatian2019task, neuhaus2020task} that the pre-quantization processing which is optimal with vector quantizers, i.e., recovery of the \ac{mmse} estimate of $\myVec{s}$ from $\myVec{x}$, is no longer optimal when using scalar quantization, and that characterizing the optimal pre-quantization processing in such cases is very difficult in general. 

While vector quantizers allow to achieve more accurate digital representations of the acquired analog signal compared to their scalar counterparts \cite[Ch. 23]{polyanskiy2015lecture}, one is often interested in utilizing scalar quantizers. For instance, when quantization is part of an acquisition procedure involving \acp{adc}, as \acp{adc} often apply the same continuous-to-discrete mapping to each sample, which is most commonly based on a uniform partition of the real line, i.e., scalar uniform quantization \cite{walden1999analog}. Nonetheless, in the presence of a task, one is not interested in recovering the analog signal, but rather estimate some underlying information embedded in it. This motivates the analysis of how to incorporate the presence of a task in the design of a quantization system utilizing scalar quantizers, and whether the distortion induced by conventional scalar quantization can be mitigated when recovering the task.

\subsubsection{System Model}
\begin{figure}
	\centering
	\includegraphics[width=6cm]{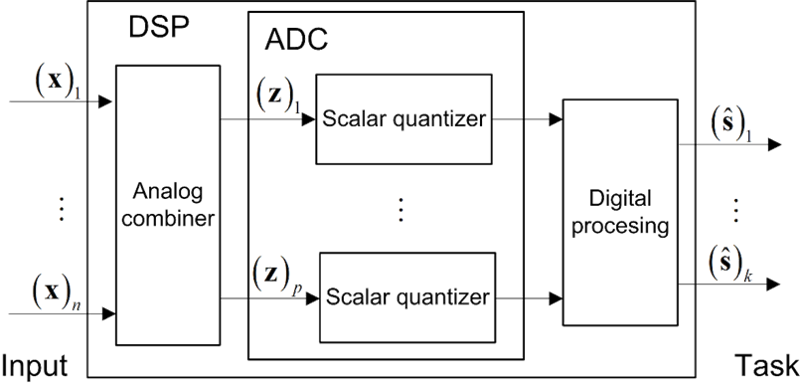} 
	\vspace{-0.2cm}
	\caption{Block diagram of considered goal-oriented quantization systems.} 
	\label{fig:HybridSystem2}
\end{figure}

Consider a goal-oriented quantization system comprised of pre-quantization mapping (termed henceforth analog combining\footnote{we use the term {\em analog} to clearly distinguish the signal prior to its quantization, complying with applications where such systems are employed to acquire analog signals using \acp{adc}. Nonetheless, the resulting formulation still holds when compressing digital data in a goal-oriented manner.}) and quantization rule. An example for such a system  is illustrated in Fig. \ref{fig:HybridSystem2}.  Here, we focus on a {\em model-based setting}~\cite{shlezinger2023model}, where the system components are to be be jointly optimized based on knowledge of the conditional distribution relating the input $\myX$ to the task vector $\myS$, denoted $P_{\myX | \myS}$.

	In order to obtain a meaningful and tractable characterization of the goal-oriented quantization  system , we henceforth introduce two model assumptions upon which we base our results in the remainder of this section: 
\begin{enumerate}[label={\em A\arabic*}]
    \item \label{itm:A1} We consider the problem of estimating the task $\myS$ in the \ac{mse} sense, namely, our performance measure is the \ac{mse} $\E\{\|\myS - \hat{\myS}\|^2 \} $.
    \item \label{itm:A2} We focus on uniform quantizers, and model the their operation in our derivations as non-subtractive uniform dithered quantizers \cite{gray1993dithered}. 
\end{enumerate}

	Model assumption \ref{itm:A1} implies that the fidelity of an estimate $\hat{\myS}$ can be represented as a sum of the \ac{mmse} and the excess \ac{mse} with  respect to the \ac{mmse} estimate $\tilde{\myS} = \E\{\myS|\myX\}$, as $\E\{\|\myS - \hat{\myS}\|^2 \} = \E\{\|\myS - \tilde{\myS} \|^2 \} + \E\{\|\tilde{\myS}  - \hat{\myS}\|^2 \}$. Consequently, in the following we characterize the performance in terms of the excess \ac{mse} $ \E\{\|\tilde{\myS}  - \hat{\myS}\|^2 \}$. Since $\tilde{\myS}$ is a function of $\myX$, we divide our analysis based on the nature of this function, considering first linear functions, extending to quadratic and more general forms.  
	
Model assumption \ref{itm:A2} imposes a structure on the scalar quantization mapping. 
To formulate the resulting input-output relationship of the quantizers, let $\DynRange$ denote the support of the quantizer, and define $\DynInt \triangleq \frac{2\DynRange}{\TilM}$ as the quantization spacing. 
The output of the uniform \ac{adc} with input sequence $z_1, z_2, \ldots, z_\lenZ$ can be written as $Q\left( z_i\right)  =  q\left(z_i + u_i \right) $, where $u_1, u_2, \ldots, u_\lenZ$ are i.i.d. \acp{rv} uniformly distributed over $\left[-\frac{\DynInt}{2},\frac{\DynInt}{2} \right]$, mutually independent of the input, representing the dither signal.
The function $q(\cdot)$, which implements the uniform quantization, is given by  
\begin{equation}
q(z) = \begin{cases}
-\DynRange + \DynInt\left(l - \frac{1}{2} \right)   & \begin{array}{c}
     z - l  \DynInt \in  \left[-\frac{\DynInt}{2},\frac{\DynInt}{2} \right],  \\
     l \in \{0,1, \ldots, \TilM - 1 \} 
\end{array}
   \\
{\rm sign}\left(z\right) \left( \DynRange - \frac{\DynInt}{2}\right)    & |z| > \DynRange.
\end{cases}
\label{eqn:UniQuant} 
\end{equation}   
When $\TilM = 2$, the resulting quantizer is a standard one-bit sign quantizer of the form $q(z) = c \cdot {\rm sign}(z)$, where $c >0$ is determined by the support $\DynRange$.

	Dithered quantizers significantly facilitate the analysis, due to the  following favorable property: When operating within the support, the output can be written as the sum of the input and an additive zero-mean white quantization noise signal uncorrelated with the input. The drawback of adding dither is that it increases the energy of the quantization noise, namely, it results in increased distortion \cite{gray1993dithered}. Nonetheless, the favorable property of dithered quantization is also satisfied in uniform quantization  {\em without dithering} for inputs with bandlimited characteristic functions, and is approximately satisfied for various families of input distributions \cite{widrow1996statistical}. Consequently, while our  analysis assumes dithered quantization, exploiting  the resulting statistical properties of the quantization noise, the proposed system is applicable without dithering, as we demonstrate in our numerical study. 

\subsubsection{Linear Estimation Tasks}
We begin by focusing on scenarios in which the stochastic relationship between the vector of interest $\myS$ and the observations $\myX$ are such that the \ac{mmse} estimate of $\myS$ from $\myX$  is a linear function of $\myX$, i.e., $\exists \LmmseMat \in \mySet{R}^{\lenS \times \lenX}$ such that $\tilde{\myS} = \LmmseMat \myX$.  	
Accordingly,  we restrict the analog combining and the digital mapping components in Fig. \ref{fig:HybridSystem2} to be linear, namely, $\myZ = \myA \myX$ and $\hat{\myS} = \myB Q(\myZ)$, for some $\myA \in \mySet{R}^{\lenZ \times \lenX}$ and  $\myB \in \mySet{R}^{\lenS \times \lenZ}$. 
%
%
%
%
By focusing on these setups, we are able to explicitly derive the achievable distortion and to characterize the system which minimizes the \ac{mse}. This derivation reveals some non-trivial insights. For example, we show that the optimal approach when using vector quantizers, namely, to quantize the \ac{mmse} estimate \cite{wolf1970transmission}, is no longer optimal when using  standard scalar quantizers. Furthermore, as detailed in the sequel, our analysis provides  guidelines for designing  goal-oriented quantization systems which can be used for more general relationships between $\myS$ and $\myX$, such as the recovery of quadratic tasks.

Let $\CovMat{\myX}$ be the covariance matrix of $\myX$, assumed to be non-singular.
Before we study the overall goal-oriented quantization system, we first derive the  digital processing matrix which minimizes the \ac{mse} for a given analog combiner $\myA$ and the resulting  \ac{mse},  stated in the following lemma \cite[Lem. 1]{shlezinger2018hardware}:
\begin{lemma}
	\label{lem:ThmProof1}
	\begin{subequations}
	For any analog combining matrix $\myA$ and support $\DynRange$ such that the quantizers operate within their support, i.e., $\Pr \left( \big|\left( \myA \myX\right)_l + u_l\big| > \DynRange \right) = 0$,   
	the  digital processing matrix which minimizes the \ac{mse} is given by 
	\begin{equation}
	\myB\opt \left( \myA\right)  = \LmmseMat\CovMat{\myX}\myA^T\bigg( \myA\CovMat{\myX}\myA^T + \frac{{2{\DynRange^2}}}{{3\TilM^2}}{\myI_\lenZ} \bigg)^{ - 1},
	\end{equation}
	and the achievable excess \ac{mse}, denoted ${\rm MSE}\left(\myA\right) 
	= \mathop {\min }_{\myB} \E\big\{ \big\| \tilde{\myS}-\hat{\myS} \big\|^2 \big\} $, is 
	\begin{align}
	{\rm MSE}\left(\myA\right) 
	\!=\! 
	{\rm Tr} \bigg(& \LmmseMat \CovMat{\myX}\LmmseMat^T\! -\! \LmmseMat\CovMat{\myX}\myA^T\!\bigg( \myA\CovMat{\myX}\myA^T\! \notag \\ 
 &+ \!\frac{{2{\DynRange^2}}}{{3\TilM^2}}{{\bf{I}}_\lenZ} \bigg)^{ - 1}\!\!\!\!\!\myA\CovMat{\myX}\LmmseMat^T \bigg).
	\end{align} 
	\end{subequations}	
\end{lemma}

The digital processing matrix in Lemma~\ref{lem:ThmProof1} is the linear  \ac{mmse} estimator of $\myS$ from the vector $\myA\myX + \myVec{e}$, where $\myVec{e}$ represents the quantization noise, which is white and uncorrelated with $\myA\myX$. This stochastic representation is a result of the usage of non-overloaded dithered quantizers. 
Nonetheless, in the following we use the model on which Lemma~\ref{lem:ThmProof1} is based  to design goal-oriented quantizers operating with small yet non-zero probability of overloading, i.e.,  $\Pr \left( \big|\left( \myA \myX\right)_l + u_l\big| > \DynRange \right) \approx 0$ for each $l$. In such cases modeling  $\myA\myX$ and $\myVec{e}$ as uncorrelated becomes a reliable approximation. Therefore, in order to use Lemma \ref{lem:ThmProof1} to design goal-oriented quantizers,  we explicitly require to avoid overloading with high probability. This is achieved by fixing $\DynRange$ to be some multiple $\myEta$ of the maximal standard deviation of the input, allowing to bound the overload probability via  Chebyshev's inequality \cite[Pg. 64]{cover2012elements}. 

We now use Lemma \ref{lem:ThmProof1} to obtain the analog combining matrix $\myA\opt$ which minimizes the \ac{mse} and the resulting system.
Define the matrix $\tilde{\LmmseMat} \triangleq \LmmseMat \CovMat{\myX}^{1/2}$,  let $\{ \eig{\tilde{\LmmseMat},i} \}$ be its singular values arranged in a descending order, and set  $\MyKappa \triangleq \myEta^2 \big(1 - \frac{ \myEta^2 }{3\TilM^2}\big)^{-1}$. Note that for $i > {\rm rank} \big( \tilde{\LmmseMat}\big)$, $\eig{\tilde{\LmmseMat},i}  = 0$. The resulting goal-oriented quantization system is stated in the following theorem \cite[Thm. 1]{shlezinger2018hardware}:
\begin{theorem}
	\label{thm:OptimalDes}
	\begin{subequations}
		\label{eqn:OptimalDes}
		For the goal-oriented quantization system under linear estimation tasks, the analog combining matrix $\myA\opt$ is given by $\myA\opt = \myMat{U}_{\myA} \myMat{\Lambda}_{\myA} \myMat{V}_{\myA}^T \CovMat{\myX}^{-1/2}$, where  $\myMat{V}_{\myA} \in \mySet{R}^{\lenX \times \lenX}$ is the right singular vectors matrix of  $\tilde{\LmmseMat}$; 
		  $\myMat{\Lambda}_{\myA} \in \mySet{R}^{\lenZ \times \lenX}$ is a diagonal matrix with diagonal entries  
			\begin{equation}
			\label{eqn:OptimalDesA}
			\left( \myMat{\Lambda}_{\myA}\right)_{i,i}^2 = \frac{{2{\MyKappa }}}{{3\TilM^2} \cdot \lenZ}\left( {\Wlevel  \cdot\eig{\tilde{\LmmseMat},i} - 1} \right)^ +,
			\end{equation}
			with  $\Wlevel$  set such that $\frac{{2{\MyKappa }}}{{3\TilM^2} \cdot \lenZ}\sum_{i=1}^{\lenZ} \big( {\Wlevel  \cdot\eig{\tilde{\LmmseMat},i} - 1} \big)^ + = 1$; 
			and $\myMat{U}_{\myA} \in \mySet{R}^{\lenZ \times \lenZ}$ is a unitary matrix which guarantees that  
			 $\myMat{U}_{\myA}\myMat{\Lambda}_{\myA}\myMat{\Lambda}_{\myA}^T\myMat{U}_{\myA}^T$ is weakly majorized by all possible rotations of  $\myMat{\Lambda}_{\myA}\myMat{\Lambda}_{\myA}^T$. 
		The support of the \ac{adc} is given by $	\DynRange^2   =   \frac{ \MyKappa}{\lenZ}$, 
		and the digital processing matrix is equal to	
		\begin{equation}
		\label{eqn:OptimalDesB}
		\myB\opt\left( \myA\opt\right) = \tilde{\LmmseMat} \myMat{V}_{\myA}\myMat{\Lambda}_{\myA}^T\left( \myMat{\Lambda}_{\myA} \myMat{\Lambda}_{\myA}^T+ \frac{{2{\DynRange^2}}}{{3\TilM^2}}\myI_{\lenZ} \right)^{ - 1}\!\! \myMat{U}_{\myA}^T.
		\end{equation}
		The resulting minimal achievable excess \ac{mse} is
		\begin{equation}
		\label{eqn:OptimalDesMSE}	
		\E \left\{\left\|\tilde{\myS} \!-\! \hat{\myS} \right\|^2  \right\}\! = \!
		\begin{cases} 
		\sum\limits_{i=1}^{\lenS}   \frac{ \eig{\tilde{\LmmseMat},i}^2} {\left(\Wlevel \cdot\eig{\tilde{\LmmseMat},i} -  1 \right)^+ \!+ \!1}, &\lenZ \!\ge\! \lenS \\
		\sum\limits_{i=1}^{\lenZ}   \frac{ \eig{\tilde{\LmmseMat},i}^2} {\left(\Wlevel \cdot\eig{\tilde{\LmmseMat},i} -  1 \right)^+ \!+\! 1} \!+\! \sum\limits_{i\!=\!\lenZ\!+\!1}^{\lenS}  \eig{\tilde{\LmmseMat},i}^2, & \lenZ\! <\! \lenS.
		\end{cases} 
		\end{equation}
	\end{subequations}
\end{theorem}  

The majorizing unitary matrix $\myMat{U}_{\myA}$ is guaranteed to exist by  \cite[Cor. 2.1]{palomar2007mimo}, and can obtained via, e.g., \cite[Alg. 2.2]{palomar2007mimo}. 
Since the design objective is the \ac{mse} by \ref{itm:A1}, the optimal quantization system utilizing vector quantizers is known to recover $\tilde{\myS} = \LmmseMat\myX$ in the analog domain \cite{wolf1970transmission}. In the presence  of scalar quantizers, Theorem \ref{thm:OptimalDes} reveals two main differences in the desired pre-quantization mapping: First, the analog combiner essentially nullifies the weak eigenmodes of the correlation matrix of the \ac{mmse} estimate in \eqref{eqn:OptimalDesA}, as these eigenmodes are likely to become indistinguishable by finite resolution uniform scalar quantization. 
Then, the unitary rotation matrix  $\myMat{U}_{\myA}$, which guarantees that the entries of $\myZ$ have the same variance, minimizes the maximal variance of the quantized variables, allowing to use relatively fine quantization at a given resolution without risking high overloading probability. This combined operation of the analog mapping trades estimation error and quantization accuracy, allowing to optimize the digital representation in light of the task. 


The characterization of the goal-oriented quantization system in Theorem~\ref{thm:OptimalDes} gives rise to the following non-trivial insights: 
$1)$ 	In order to minimize the \ac{mse}, $\lenZ$ must not be larger than the rank of the covariance matrix of  $\tilde{\myS}$ \cite[Cor. 1]{shlezinger2018hardware}. This implies that reducing the dimensionality of the input prior to quantization contributes to recovering the task vector as higher resolution quantizers can be used without violating the overall bit constraint;
and $2)$ 	When the covariance matrix of $\tilde{\myS}$ is non-singular, quantizing the \ac{mmse} estimate minimizes the \ac{mse} if and only if the covariance matrix of  $\tilde{\myS}$ equals   $\myI_{\lenS}$ up to a constant factor \cite[Cor. 4]{shlezinger2018hardware}. 
This indicates that, except for very specific statistical models, quantizing the entries of the \ac{mmse} estimate vector, which is the optimal strategy when using vector quantizers \cite{wolf1970transmission}, does not minimize the \ac{mse} when using uniform scalar quantizers.


\subsubsection{Quadratic Estimation Tasks}
So far, we showed that allowing the analog mapping to reduce  dimensionality and rotate the quantized signal can contribute to the overall recovery performance by balancing estimation and quantization errors. However, this analysis was carried out only for scenarios in which $\tilde{\myS}$ is a linear function of $\myX$, resulting in   $\E\big\{\myS | \myZ \big\}$ being a linear function of the input to the quantizers $\myZ$.  In many scenarios of interest, such as covariance estimation \cite{rodrigues2017rate} and \ac{doa} recovery \cite{yu2016doa} from quantized measurements, the desired information can be extracted from a quadratic function of the measurements, i.e., functions $\{\myX^T \myMat{C}_{i} \myX\}_{i=1}^{\lenS}$, where each $\myMat{C}_{i} \in \mySet{R}^{\lenX \times \lenX}$ is symmetric. 
	
	Here, we show how the analysis of the previous section can be applied for designing goal-oriented quantizers for the task of recovering non-linear functions of $\myX$ under quantization constraints, focusing on quadratic functions and Gaussian inputs. 
%
%
Our strategy is based on identifying a family of analog mappings $h(\cdot)$ for which $\myZ$ corresponds to the linear task scenario. To that aim, we use \ac{pic}-based analysis \cite{du2017principal}, which provides a decomposition of the statistical relationship between two \acp{rv}, that is directly related to \ac{mmse} estimation. 
In particular, for a pair of \acp{rv} $(x,y)$, the principal inertia functions $\{f_i(\cdot)\}$ and $\{g_i(\cdot)\}$ form an orthonormal basis spanning the Hilbert space of functions of $x$ and $y$, respectively, which diagonalize \ac{mmse} estimation, i.e., there exists a set of scalar coefficients $\{\rho_i\}$ such that $\E\{f_i(x)|y\} = \rho_i g+i(y)$ and $\E\{g_i(y)|x\} = \rho_i f_i(x)$. The benefit of using \acp{pic} in our context is their ability to decompose functions of the observations in a manner which reflects on the structure of the \ac{mmse} estimate. In particular, here we use this tool to identify a transformation of the input $\myX$ under which recovering quadratic functions of it is converted to a linear manipulation.
Defining $\bar{\myX} \triangleq {\rm vec}(\myX \myX^T)$, this results in the following theorem \cite[Thm. 1]{salamatian2019task}:
\begin{theorem}
	\label{thm:PICQuad}
	For any $\lenZ \times \lenX^2$ matrix $\myMat{A}$ with $\lenZ \le \lenX^2$, the \ac{mmse} estimate of $f(\myX) = \myX^T\myMat{C}\myX$ from  the vector $\myZ = \myMat{A}(\bar{\myX} - \E\{\bar{\myX}\})$ can be written as
	\begin{equation}
	\label{eqn:PICQuad}
	\E\big\{f(\myX) | \myZ  \big\} = \myVec{d}^T \myZ + \E\{f(\myX)\},
	\end{equation}
	for some $\lenZ \times 1$ vector $\myVec{d}$, which depends on $\myMat{C}$, $\myMat{A}$, and the covariance of $\myX$. 
\end{theorem}

Theorem \ref{thm:PICQuad} implies that the goal-oriented quantization system design guidelines proposed in Theorem \ref{thm:OptimalDes} can be utilized to facilitate the recovery of quadratic functions from quantized measurements by applying analog mappings of the form $\myZ = \myMat{A}h(\myX) = \myMat{A}(\bar{\myX} - \E\{\bar{\myX}\})$. Here the matrix $\myMat{A} \in \mySet{R}^{\lenZ \times \lenX^2}$ encapsulates the ability to reduce  the dimensionality and to rotate the quantized vector, and can be designed via Theorem \ref{thm:OptimalDes} by replacing the input $\myX$ with $\bar{\myX} - \E\{\bar{\myX}\}$.   

Although Theorem~\ref{thm:PICQuad} specifically considers functionals $f(\myX)$ of a quadratic form,   analogous schemes could be constructed for broader classes of functions. The main feature of Theorem~\ref{thm:PICQuad} is the ability to represent $\E\{f(\myX)|\myZ\}$ either exactly, or possibly approximately, as a linear function of $\myZ = \myA h(\myX)$ for some transformation $h(\cdot)$. Once the analog mapping satisfies this request, Theorem \ref{thm:OptimalDes} can be applied to optimize the overall recovery accuracy of the quantization system. 
Formulated in terms of \acp{pic}, the choice of  $h(\cdot)$ imposes  structure on the joint distribution $(\myX,\myZ)$. Consequently, when the task is to recover a function $f(\myX)$ which can be decomposed using \acp{pic} as $f(\myX) = \sum \alpha_i f_i(\myX)$, any analog processing, which results in $\myZ$ such that 
\begin{align}
\E\{f(\myX)|\myZ\} \approx \sum_{i = 1}^l \alpha_i \rho_i (\myZ)_i + \E\{f(\myX)\},  
\end{align}
 would allow to design the analog pre-quantization step using existing tools derived for setups in which the \ac{mmse} estimate is linear. This implies that when recovering  some function $f(\myX)$, the structure of the analog mapping should be designed as to yield linear basis functions $g_i(\myZ)$, allowing the resulting system to be optimized using Theorem \ref{thm:OptimalDes}.


\subsection{Rate-Distortion-Perception Theory}
\label{subsec:rdp}

Rate-distortion-perception theory is a recent development at the intersection of information theory and machine learning that aims to characterize the fundamental limits of perception-aware image compression and provide the architectural principles for related neural network model design.
To facilitate a deeper understanding of the innovative elements of this new theory, we will begin by providing a concise overview of classical rate-distortion theory. Consider a length-$n$ lossy source coding system consisting of a stochastic encoder and a stochastic decoder that share a random seed $\omega$ distributed over $\Omega$. The encoder maps the given source sequence $\myX\in\mathcal{X}^n$ along with $\omega$ to an index $m\in\mathcal{M}$ according to certain conditional distribution $P_{m|\myX\omega}$ while the decoder generates a reconstruction sequence $\hat{\myX}\in\hat{\mathcal{X}}^n$ based on $m$ and $\omega$ according to some conditional distribution $P_{\hat{\myX}|m\omega}$. It is assumed that $\myX$ and $\omega$ are independent. Let $d:\mathcal{X}^n\times\hat{\mathcal{X}}^n\rightarrow\mathcal{R}^+$ be a distortion measure. We say distortion level $D$ is achievable  for source distribution $P_{\myX}$ at blocklength $n$ subject to compression rate constraint $R$ and common randomness rate constraint $R_c$ if there exists a length-$n$ lossy source coding system such that
\begin{align}
	\frac{1}{n}\mathbb{E}\{d(\myX,\hat{\myX})\}\leq D,\hspace{0.05in}
	\frac{1}{n}\log |\mathcal{M}|\leq R,\hspace{0.05in}
	\frac{1}{n}\log |\Omega|\leq R_c.
\end{align}
The infimum of such achievable $D$ is denoted  $D^{(n)}(R,R_c)$. It is easy to show that the shared random seed can be eliminated without affecting the minimum achievable distortion, namely, $D^{(n)}(R,0)=D^{(n)}(R,\infty)$.
Therefore, we shall simply denote $D^{(n)}(R,R_c)$ by $D^{(n)}(R)$. In fact, not only is common randomness unnecessary, but both the encoder and decoder can also be restricted to deterministic functions without incurring a performance penalty. As such, designing a length-$n$ lossy source coding system boils down to design a $Q^{n,n}_M(\cdot)$ quantizer (see Section \ref{ssec:GOQaunt}).

The central result of rate-distortion theory is the single-letter characterization of the performance limit of lossy source coding in the asymptotic setting $n\rightarrow\infty$. Specifically, the celebrated rate-distortion theorem \cite{cover2012elements} states that if $\myX$ is an sequence of i.i.d.  \acp{rv} with marginal distribution $P_{x}$ and the distortion measure is of the additive form $d(\myX,\hat{\myX})=\sum_{i=1}^n\delta(x_i,\hat{x}_i)$, 
then
\begin{align}	\lim\limits_{n\rightarrow\infty}D^{(n)}(R)=\Delta(R),\label{eq:RD}
\end{align}
where $\Delta(R)$ is the distortion-rate function: 
\begin{align}	\Delta(R)\triangleq&\inf\limits_{P_{\hat{x}|x}}\mathbb{E}\{\delta(x,\hat{x})\}\\
	\mbox{subject to}&\quad I(x;\hat{x})\leq R.
\end{align}

It has been long recognized that the fidelity of a compressed image with respect to the original version does not necesarily reflect its realism. For example, \ac{mse}-based compression often leads to blurred images. As a remedy, a more comprehensive evaluation is used to guide image compression by complementing the distortion measure $d$ with a perception measure $\phi$. In practice, $\phi$ is typically chosen to be a divergence. The rationale behind this choice is that images sampled from the same distribution are of similar perceptual qualities and consequently the perception loss can be quantified by the divergence between pre- and post-compression image distributions.
The advent of perception-aware lossy source coding naturally calls for a corresponding mathematical framework, and rate-distortion-perception theory is developed against this backdrop. Now we proceed to  provide 
an exposition of this new theory with the focus placed on the architectural principles and computable performance limits. 

Length-$n$ perception-aware lossy source coding generalizes its perception-agnostic counterpart by imposing an extra perception constraint\footnote{See \cite{chen2022rate, sadafconditional2024,xie2024} for some
alternative formulations of the perception constraint.}
\begin{align}
	\frac{1}{n}\phi(P_{\myX},P_{\hat{\myX}})\leq P,\label{eq:perception_oneshot}
\end{align}
and the corresponding minimum achievable distortion is denoted  	$D^{(n)}(R,R_c,P)$. Here $\phi$ is a divergence in the sense that $\phi(P_{\myX},P_{\hat{\myX}})\geq 0$ for all $(P_{\myX},P_{\hat{\myX}})$ and the equality holds if and only if $P_{\myX}=P_{\hat{\myX}}$. Note that conventional lossy source coding can be viewed as a degenerate form of perception-aware lossy source coding with $P=\infty$. 

A generic perception-aware lossy source coding scheme can be roughly described as follows: 
\begin{itemize}
\item {\em Codebook construction:} Construct $|\Omega|\approx 2^{nR_c}$ codebooks, each with $|\mathcal{M}|\approx 2^{nR}$ codewords.

\item {\em Encoding:} By viewing each codebook as a lossy source code for $P_{\myX}$, the encoder performs lossy source encoding on $\myX$ using the codebook specified by the random seed $\omega$ and sends the codeword index $m$ to the decoder.

\item {\em Decoding:} Given $(m,\omega)$, the decoder recovers the codeword selected by the encoder. By viewing all $|\Omega|$ codebooks collectively as a lossy source code for  $P_{\hat{\myX}}$ and  $m$ as the codeword index resulted from encoding $\hat{\myX}$ using this lossy source code, the decoder generates $\hat{\myX}$ based on the recovered codeword by performing the stochastic inverse of this fictitious encoding operation.
\end{itemize}
In contrast to conventional lossy source coding, the decoder in perception-aware lossy source coding is not necessarily deterministic. This is because the encoder output $m$ and the shared random seed $\omega$ might not supply enough randomness to produce a reconstruction distribution $P_{\hat{\myX}}$  that can meet the perception constraint. 
However, with the increase of the amount of common randomness, the optimal perception-aware lossy source coding system at a given compression rate $R$ will gradually reduce 
the generative functionality of the decoder and in the end make $\hat{\myX}$  fully determined by $(\myX, \omega)$. Indeed, when $R_c$ is sufficiently large, it is possible to construct the $|\Omega|$ codebooks in such a way that they jointly form a lossless source code for $P_{\hat{\myX}}$.


In addition to the degenerate case $P=\infty$, the current knowledge regarding $D^{(n)}(R,R_c,P)$
is largely restricted to the case with no common randomness (i.e., $R_c=0$),  the case with unlimited common randomness (i.e., $R_c=\infty$), and the case with perfect realism (i.e., $P=0$). For ease of exposition, we restrict $d$ and $\phi$ to the form $d(\myX,\hat{\myX})=\sum_{i=1}^n\delta(x_i,\hat{x}_i)$ 
and  
\begin{align}	\phi(P_{\myX},P_{\hat{\myX}})=\inf\limits_{P_{\myX\hat{\myX}}\in\Pi(P_{\myX},P_{\hat{\myX}})}\sum\limits_{i=1}^n\mathbb{E}\{c(x_i,\hat{x}_i)\},\label{eq:specialform}
\end{align}
where   $c(x,\hat{x})\geq 0$ with equality if and only if $x=\hat{x}$, and $\Pi(P_{\myX},P_{\hat{\myX}})$ is the set of couplings of $P_{\tilde{\myX}}$ and $P_{\hat{\myX}}$. Choosing $c(x,\hat{x})=\|x-\hat{x}\|^2$ in (\ref{eq:specialform}) yields the squared Wasserstein-2 distance $W^2_2(P_{\myX},P_{\hat{\myX}})$.

\subsubsection{No Common Randomness} Here we shall exclusively focus on the \ac{mse} distortion measure 
(i.e.,  $\delta(x,\hat{x})=\|x-\hat{x}\|^2$).  Under this distortion measure, 
the decoder should first produce $\tilde{\myX}\triangleq\mathbb{E}\{\myX|m\}$, namely, the \ac{mmse} estimate $\tilde{\myX}$ of the source sequence $\myX$ based on the received index $m$, then use it to generate the reconstruction sequence $\hat{\myX}$ through the transportation plan $P_{\hat{\myX}|\tilde{\myX}}$ that attains $W^2_2(P_{\tilde{\myX}},P_{\hat{\myX}})$, where $P_{\hat{\myX}}$ is chosen such that (\ref{eq:perception_oneshot}) is satisfied and $\mathbb{E}\{\|\myX-\tilde{\myX}\|^2\}+W^2_2(P_{\tilde{\myX}},P_{\hat{\myX}})$ is minimized \cite[Thm. 4]{zhang2021universal}.


Further results can be obtained when the perception measure is given by the squared Wasserstein-2 distance (i.e., $c(x,\hat{x})=\|x-\hat{x}\|^2$). Given any encoder satisfying the rate constraint $R$, 
we can use the \ac{mmse} estimate $\tilde{\myX}$ induced by its output $m$ to generate $\tilde{\myX}'$ such that $P_{\tilde{\myX}'}=P_{\myX}$ and $\mathbb{E}\{\|\tilde{\myX}'-\tilde{\myX}\|^2\}=W^2_2(P_{\myX},P_{\tilde{\myX}})$. Clearly, 
$\mathbb{E}\{\|\myX-\tilde{\myX}'\|^2\}=\mathbb{E}\{\|\myX-\tilde{\myX}\|^2\}+W^2_2(P_{\myX},P_{\tilde{\myX}})$ and $W^2_2(P_{\myX},P_{\tilde{\myX}'})=0$.
Note that $\tilde{\myX}$ and $\tilde{\myX}'$ are the two extreme reconstructions based on $m$: the former has the minimum distortion loss while the other has the minimum perception loss. Interpolating $\tilde{\myX}$ and $\tilde{\myX}'$ gives $\hat{\myX}(\alpha):=\alpha\tilde{\myX}+(1-\alpha)\tilde{\myX}'$, $\alpha\in[0,1]$.
It can be verified that
$\mathbb{E}\{\|\myX-\hat{\myX}(\alpha)\|^2\}=\mathbb{E}\{\|\myX-\tilde{\myX}\|^2\}+(1-\alpha)^2W^2_2(P_{\myX},P_{\tilde{\myX}})$ and $W^2_2(P_{\myX},P_{\hat{\myX}(\alpha)})=\alpha^2W^2_2(P_{\myX},P_{\tilde{\myX}})$.
Setting $\frac{1}{n}W^2_2(P_{\myX},P_{\hat{\myX}(\alpha)})=P$ yields
\begin{align}
	&\frac{1}{n}\mathbb{E}\{\|\myX-\hat{\myX}(\alpha)\|^2\}\nonumber\\
	&=\frac{1}{n}\mathbb{E}\{\|\myX-\tilde{\myX}\|^2\}+[(\frac{1}{\sqrt{n}}W_2(P_{\myX},P_{\tilde{\myX}})-\sqrt{P})_+]^2,\label{eq:DPtradeoff}
\end{align}
where $(\cdot)_+\triangleq\max\{\cdot,0\}$
It turns out that this is the minimum distortion achievable at perception loss $P$ based on the output of this encoder \cite[Thm. 1]{Freirich2021}. Moreover, $\mathbb{E}\{\|\myX-\tilde{\myX}\|^2\}$ and $W^2_2(P_{\myX},P_{\tilde{\myX}})$ are simultaneously minimized when the encoder is taken from the  optimal $Q^{n,n}_M(\cdot)$ quantizer for source distribution $P_{\myX}$ at rate $R$. For such an encoder,  we have  \cite{yan2022controllable}
\begin{align}
	\frac{1}{n}\mathbb{E}\{\|\myX-\tilde{\myX}\|^2\}=\frac{1}{n}W^2_2(P_{\myX},P_{\hat{\myX}})=D^{(n)}(R).\label{eq:minDP}
\end{align}
Substituting (\ref{eq:minDP}) into (\ref{eq:DPtradeoff}) shows
\begin{align}
	D^{(n)}(R,0,P)=D^{(n)}(R)+[(\sqrt{D^{(n)}(R)}-\sqrt{P})_+]^2,\label{eq:oneshotL2W2}
\end{align}
which, together with (\ref{eq:RD}), further implies that when $\myX$ is a sequence of i.i.d. \acp{rv} with marginal distribution $P_{x}$,
\begin{align}
\lim\limits_{n\rightarrow\infty}D^{(n)}(R,0,P)=\Delta(R)+[(\sqrt{\Delta(R)}-\sqrt{P})_+]^2.\label{eq:Rc=0}
\end{align}

The above analysis indicates that under \ac{mse} distortion measure and squared Wasserstein-2 perception measure, one can simply set the encoder  to be the same as that in the optimal quantizer with distortion $D^{(n)}(R)$
 whereas the decoder needs to be designed according to the perception constraint $P$. When $P\geq D^{(n)}(R)$, the decoder degenerates to the \ac{mmse} estimator and the end-to-end distoriton is $D^{(n)}(R)$. When $P=0$, the decoder is a generator that simulates 
 the stochastic inverse of the encoder (i.e., posterior sampling) and the end-to-end distortion is $2D^{(n)}(R)$ \cite[Thm. 2]{yan2021perceptual}.
For $P\in(0,D^{(n)}(R))$, the decoder operates by properly interpolating the outputs of these two extreme modes.

The observation that the optimal encoder can be specified without the knowledge of perception constraint naturally gives rise to the notion of universal representations \cite{zhang2021universal}. While the analysis here focuses on the case of \ac{mse} distortion measure and squared Wasserstein-2 perception measure, the existence of universal representations has been observed in a much broader context. This paves the way for designing machine learning models with a fixed encoder that are approximately optimal for a multitude of operational objectives. It also open up promising prospects for task-adaptive communications without completely compromising the universal bit interface.

\subsubsection{Unlimited Common Randomness} 

The behavior of $D^{(n)}(R,\infty,P)$ is better understood in the asymptotic setting $n\rightarrow\infty$. 
Specifically, we have \cite[Thm. 3]{theis2021coding}
\begin{align}
\lim\limits_{n\rightarrow\infty}D^{(n)}(R,\infty,P)=\Delta(R,P)\label{eq:Rc=inf}
\end{align}
when $\myX$ is a sequence of i.i.d. \acp{rv} with marginal distribution $P_{x}$, where $\Delta(R,P)$ is the distortion-rate-perception function introduced by Blau and Michaeli \cite{blau2019rethinking}:
\begin{align}
	\Delta(R,P)\triangleq&\inf\limits_{P_{\hat{x}|x}}\mathbb{E}\{\delta(x,\hat{x})\}\\
	\mbox{subject to}&\quad I(x;\hat{x})\leq R,\hspace{0.05in}\phi(P_{x},P_{\hat{x}})\leq P.
\end{align}

The fact that the fundamental distortion-rate-perception tradeoff depends on the available amount of common randomness has interesting  implications to joint source-channel coding (JSCC).
It is well known that the channel capacity is typically higher with the state information available to both the transmitter and the receiver
as compared to the case where the  state information is unavailable or only known to the receiver. The shared knowledge of channel state is a particular form of common randomness, so it can be leveraged to improve the distortion-rate-perception tradeoff as well. 
This suggests a weak coupling principle that simultaneously exploits the state information as a resource in both channel coding and perception-aware lossy source coding.

\subsubsection{Perfect Realism} Note that setting $P=0$ forces $P_{\myX}=P_{\hat{\myX}}$ and the choice of perception measure is immaterial in this case. As shown by Saldi et al. \cite[Thm. 1]{saldi2015output} and rediscovered by  Wagner \cite[Thm. 3]{wagner2022rate}, when $\myX$ is a sequence of i.i.d. random variabs with marginal distribution $P_{x}$,
\begin{align}
\hspace{-0.09in}\lim\limits_{n\rightarrow}D^{(n)}(R,R_c,0)=&\inf\limits_{P_{u\hat{x}|x}}\mathbb{E}\{\delta(x,\hat{x})\}\label{eq:P=0}\\
\mbox{subject to}&\quad I(x;u)\leq R,\hspace{0.05in} I(\hat{x};u)\leq R+R_c,\\
&\quad P_{u\hat{x}|x}=P_{u|x}\Pdf{\hat{x}|u},\hspace{0.05in} P_{\hat{x}}=P_{x}.
\end{align}






One can readily verify that (\ref{eq:Rc=0}), (\ref{eq:Rc=inf}), and (\ref{eq:P=0}) are consistent. Nevertheless, it remains unknown how to unify them into a single-letter characterization of $\lim_{n\rightarrow\infty}D^{(n)}(R,R_c,P)$, even in  
the case of \ac{mse} distortion
measure and squared Wasserstein-2 perception measure.


\subsection{Compression for Estimation and Learning} 
\label{subsec:est}
Parameter estimation is one of the most  fundamental problems in statistics. Consider the classical estimation model 
\begin{equation}\label{eq:parest}
\myX_1,\myX_2,\dots,\myX_n \stackrel{i.i.d.}{\sim} P_{\bf{\theta}},
\end{equation}
where we observe $n$ samples from an underlying distribution $P_{\bf\theta}$, which is parameterized by a potentially large but finite set of
parameters ${\bf\theta}\in\Theta\subseteq\mathbb{R}^d$, and we want to estimate the unknown parameters ${\bf\theta}$ under a given distortion criterion, e.g., the \ac{mse} $d({\bf\theta},\hat{\bf\theta})=||{\bf\theta}-\hat{\bf\theta}||^2$. Many statistical tasks can be viewed as special cases of this general parametric model. For example, discrete distribution estimation can be viewed as a special case of this model with $d$ parameters to estimate by setting $P_\theta=(\theta_1,\dots,\theta_d)$, and $\Theta=\mathcal{P}_d=\{0\leq \theta_i\leq 1:\sum \theta_i=1\}$ is the $(d-1)$-dimensional probability simplex, i.e., $\theta_i$ corresponds to the probability of the $i$'th symbol in an alphabet of size $d$. As another canonical example, allowing $\myX_i\in\mathbb{R}^d$ and (i) $P_\theta=\mathcal{N}(\theta, I_d)$ or (ii) $P_\theta=\mathcal{N}(\bf{0}, \theta_d)$ corresponds to modeling the underlying distribution as a $d$-dimensional Gaussian distribution with (i) an unknown mean and identity covariance matrix (a.k.a. Gaussian mean estimation), or (ii) zero-mean and unknown covariance matrix (a.k.a. Gaussian covariance estimation). $P_\theta$ can also describe  a linear,  logistic or polynomial regression model, or a neural network model, in which case  $\theta$ corresponds to the  unknown coefficients of the model (e.g., the coefficients of a neural network). 

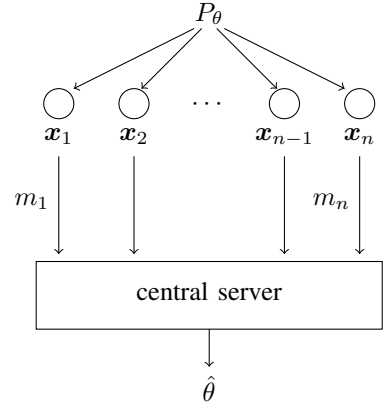
\begin{figure}\centering
        \begin{tikzpicture}
\node at (2, 1.2) {$P_\theta$}; 
\draw [->] (1.8, 1) -- (0.2, 0.2); 
\draw [->] (1.9, 1) -- (1.1, 0.2); 
\draw [->] (2.1, 1) -- (2.9, 0.2); 
\draw [->] (2.2, 1) -- (3.8, 0.2); 
\draw (0,0) circle (0.2cm); \node [below] at (0,-0.2) {$\myX_1$}; 
\draw (1,0) circle (0.2cm); \node [below] at (1,-0.2) {$\myX_2$}; 
\node at (2,0) {$\cdots$}; 
\draw (3,0) circle (0.2cm); \node [below] at (3,-0.2) {$\myX_{n-1}$}; 
\draw (4,0) circle (0.2cm); \node [below] at (4,-0.2) {$\myX_n$}; 
\draw [->] (0, -0.7) -- (0, -2); \draw [->] (1, -0.7) -- (1, -2); 
\draw [->] (3, -0.7) -- (3, -2); \draw [->] (4, -0.7) -- (4, -2);
\draw (-0.3,-3) rectangle (4.3,-2.1);  
\node at (2,-2.5) {central server}; 
\draw [->] (2,-3) -- (2, -3.5); \node [below] at (2,-3.5) {$\hat{\theta}$}; 
\node[left]  at (0, -1.3) {$m_1$};
\node [left] at (4,-1.3) {$m_n$}; 
\end{tikzpicture}
      \caption{Distributed estimation under communication constraints.\looseness=-10}\label{fig-distributed}  
\end{figure}

Unlike the traditional statistical setting where the samples $\myX_1,\dots, \myX_n$ are available to the estimator as they are, here we would like to consider the setting where each sample $\myX_i$ needs to be represented by using a fixed number of $k$ bits. More precisely, we assume that each sample $\myX_i$ needs to be encoded by a function $f_i:\mathcal{X}\rightarrow\{1,\dots,2^k\}$ that maps the observed sample $\myX_i$ to a $k$-bit  representation $m_i$. (Under this model, each sample is independently encoded into fixed $k$ bits. A variation of this model, called the sequential or interactive model, allows the encoding of the $i$'th sample to depend on all previous samples $\myX_1,\dots, \myX_{i}$.) The $k$-bit constraint can represent storage or communication constraints. For example, in a distributed setting, each sample $\myX_i$ is observed at a different node $i$ for $i=1,\dots, n$ and the compressed sample $m_i$ is communicated to a central server (see Fig.~\ref{fig-distributed}). The server aims to estimate the unknown parameters $\theta$ of the model from the messages $m_1,\dots, m_n$ under a given distortion measure.

What are optimal quantization/estimation schemes for different statistical models $P_\theta$ and distortion measures? How does the optimal estimation error depend on the bit budget $k$, dimension $d$ and the number of samples $n$? Such questions have been of significant interest in the recent literature motivated by applications where a large amount of high-dimensional data is collected by multiple nodes. Order-optimal schemes and tight information-theoretic lower bounds on the estimation error have been developed for various statistical tasks
 \cite{zhang2012communication, garg2014communication, braverman2016communication, suresh2017distributed, han2018geometric, han2018distributed, barnes2020lower, acharya2024unified}. For example, for discrete distribution estimation, \cite{han2018geometric, han2018distributed, barnes2020lower} show that 
 \begin{equation}\label{eq:minimax}
\sup_{\theta\in \mathcal{P}_d} \mathbb{E}_{P_\theta}\{\|\hat{\theta}-\theta\|_2^2\}=\frac{ c_1 d}{n\min(d,2^k)},
\end{equation}
where $c_1>0$ is a constant independent of everything else. As compared to the classical (centralized) case, where we can achieve an $\ell_2$ error of $O\left(\frac{1}{n}\right)$, this implies a factor of $d$ increase in the error when
$k$, the number of bits per sample, is small. This penalty of $d$ can be significant in many applications  of practical interest where the domain can be very large or unknown, e.g. in language modeling  $d$ can correspond to the number of all possible words a user can type, including both in and out-of-dictionary words. On the other hand, for Gaussian mean estimation where  $P_\theta=\mathcal{N}(\theta, I_d)$, the order optimal mean-squared error is given by \cite{zhang2012communication, han2018distributed, barnes2020lower}
 \begin{equation}\label{eq:minimax}
\sup_{\theta\in [-1,1]^d} \mathbb{E}_{P_\theta}\{\|\hat{\theta}-\theta\|_2^2\}=\frac{ c_2 d^2}{n\min(d,k)},
\end{equation}
revealing a linear rather than exponential dependence on $k$. Order-optimal lower bounds for both models are developed in \cite{barnes2020lower} by characterizing how $k$-bit quantization of a sample impacts the Fisher information associated with
that statistical model. These results can be extended to the corresponding distribution-free settings, empirical frequency estimation and mean estimation respectively, where each sample $\myX_i$ is now assumed to take arbitrary values in a given domain without any probabilistic assumptions on the generation of data.

This problem formulation is also relevant for distributed and federated learning \cite{mcmahan2017learning, kairouz2019advances}. In these systems, the sample $\myX_i$ can be viewed as the high-dimensional model update computed at each client node $i$, while the server aims to estimate the mean model update. Transmitting model updates between client nodes and the central node can consume substantial bandwidth, impeding the system's scalability \cite{mcmahan2017learning}. Therefore, using compression techniques to reduce bandwidth for model updates while allowing the server to accurately estimate the mean model update has become a significant research focus. Common techniques include scalar and vector quantization \cite{alistarh2017qsgd, bernstein2018signsgd}, sparsification \cite{aji2017sparse, lin2017deep, wangni2018gradient, barnes2020rtopk}, projection \cite{rothchild2020fetchsgd}, entropy encoders \cite{havasi2019minimal}, or low-rank decomposition \cite{hu2021lora}.  Furthermore, when employing a lossy compression scheme, the locally compressed model updates naturally limit the information transmitted to the central entity, thereby enhancing user privacy. If carefully designed, such compression schemes can provide or amplify differential privacy \cite{chen2020breaking, feldman2021lossless, shah2022optimal, chen2024privacy,lang2023joint}, the privacy standard widely adopted in both academia and industry. 

\section{AI-Enhanced Compression}
\label{sec:AI-Enhanced}
\subsection{Neural Goal-
Oriented Compression}
\label{subsec:Neural-Goal}
The systems considered in  Section \ref{ssec:GOQaunt} are {\em model-aware}, requiring accurate knowledge of the statistical relationship between the observations and the task. 
Two notable challenges are associated with such model-aware designs: 
\begin{enumerate}
    \item Accurate knowledge of the statistical model  may be unavailable in practice.
    \item  Even when the statistical model is perfectly known, analytically tractable characterizations  are obtained only for tasks of relatively simple form, e.g., linear and quadratic functions (as in model assumptions \ref{itm:A1}-\ref{itm:A2}).
\end{enumerate}

An alternative approach to inferring the quantization system from a statistical description, is to learn it from a set of training samples in a data-driven fashion. In particular, by utilizing machine learning methods, one can  implement goal-oriented quantizers without the need to explicitly know the underlying model and to analytically derive the proper quantization rule. Furthermore, when the parameters of the system are learned from data and not specified analytically, the quantization mapping can be optimized along with the system parameters instead of fixing a uniform scalar rule as in \eqref{eqn:UniQuant}. Finally, additional families of tasks, such as classification, can be considered by properly setting the loss function utilized in the learning process.  

\subsubsection{System Model}

Neural goal-oriented quantization operates in a data-driven manner, learning  the pre-quantization encoding, quantization mapping, and subsequent processing from a training data set. We focus here on a supervised learning setup, where the data consists of $T$ independent realizations of $\myVec{s}$ and $\myVec{x}$, denoted $\{ \myVec{s}^{(i)}, \myVec{x}^{(i)}\}_{i=1}^{t}$.

\begin{figure}
	\centering
	\includefig{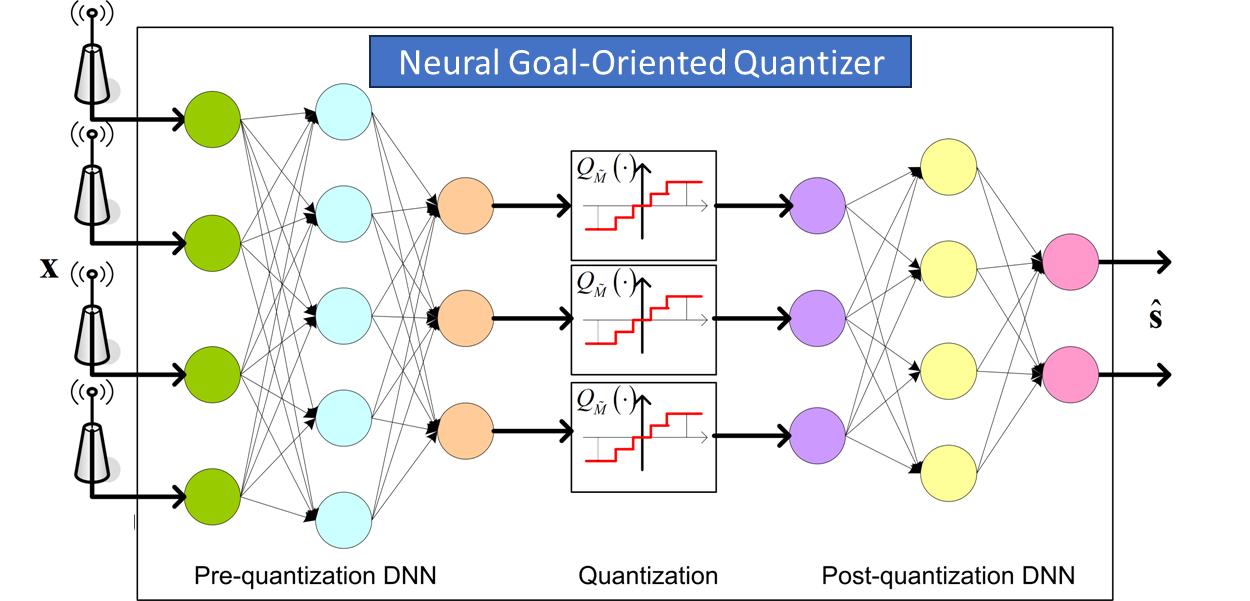}
	\caption{Neural goal-oriented quantization system with scalar quantizers illustration.} 
	\label{fig:DNNModel}
\end{figure}

Here, the  pre-quantization mapping and the  post-quantization processing are parameterized as layers of a \ac{dnn}, as illustrated in Fig.~\ref{fig:DNNModel}.  By doing so, the overall goal-oriented quantization system,  can be trained from data in an end-to-end manner using e.g., \ac{sgd} and its variant. While the proposed system focuses only on quantization, the resulting design approach can be extended to account also for sampling in addition to quantization, as considered in \cite{shlezinger2020learning}.

\subsubsection{Training Neural Goal-Oriented Quantizers}
The overall network is trained in an end-to-end manner using some variant of \ac{sgd} with back-propagation to minimize a data-based empirical risk function. The quantization layer (see Fig.~\ref{fig:DNNModel}) converts its continuous-amplitude input into a discrete quantity.  
	The non-differentiable nature of such continuous-to-discrete mappings induces a  challenge in applying \ac{sgd} for optimizing the  network parameters.  
In particular, quantization activation, which  determines the continuous regions jointly mapped into a single value, nullifies the gradient of the loss function. Thus,  straight-forward application of \ac{sgd} with back-propagation fails to properly set the pre-quantization network.
This challenge can be adressed in various manners, depending on the type of the quantizer (i.e., scalar vs. vector) and whether its mapping is learned or not.

\smallskip
{\bf Uniform Scalar Quantizers:}  
When the quantizer mapping is fixed, the core challenge is to approximate its operation during training such that one can backpropagate through it. A candidate approach to do so is to model its operation as additive noise during training~\cite{jankowski2020wireless,yang2019deep}. This formulation, which is based on an established approximate statistical modelling of uniform scalar quantizers~\cite{widrow1996statistical}, facilitates \ac{sgd}-based joint learning of the pre-quantization and post-quantization \acp{dnn}.

\smallskip 
{\bf Learned Scalar Quantizers:}
The fact that the overall goal-oriented quantization system is learned indicates that the training capabilities can also be leveraged to learn the quantization mapping itself, thus deviating from conventional uniform mappings~\cite{shlezinger2021deep,agustsson2017soft}, and also representing the operation of certain physical quantizer architectures~\cite{danial2024power}. 

This is achieved by replacing the continuous-to-discrete transformation with a non-linear activation function which has approximately the same behavior as the quantizer. Specifically, we use a sum of shifted hyperbolic tangents, which are known to closely resemble step functions in the presence of large magnitude inputs. The resulting scalar quantization mapping is given by: 
\begin{equation}
\label{eqn:tanh}
Q_{\tilde{M}}(z)=\sum_{i=1}^{\tilde{M}-1}a_{i}\tanh\left(c_{i}\cdot   z-b_{i} \right), 
\end{equation}
where $\{a_i, b_i, c_i \}$ are  real-valued parameters. When the parameters $\{c_i\}$ increase, the corresponding hyperbolic tangents approach step functions. 

In addition to learning the weights of the analog and digital \acp{dnn}, this  approach allows  to learn the quantization  function, and particularly, the best suitable constants $\{a_{i}\}$ and $\{b_{i}\}$. These tunable parameters are later used to determine the decision regions of the scalar quantizer, where the set $\{b_{i}\}$ is used for the decision regions limits while  $\{a_{i}\}$ determines the corresponding discrete values assigned to each decision region.  
The parameters $\{c_i\}$, which essentially control the resemblance of \eqref{eqn:tanh} to an actual continuous-to-discrete mapping, do not reflect on the quantization rule, and are thus not learned from training.

\smallskip 
{\bf Learned Vector Quantizers:} 
The above approach for learning quantization mappings along with the end-to-end goal-oriented system is geared towards scalar quantizers (where the parameters dictate the edges of the decision regions and their associated representations). 
Such an approach does not naturally extend to learned vector quantization. 
An alternative mechanism, leveraging the lower distortion caused by vector quantizers than scalar quantizers, is based on the established  \ac{vqvae}   \cite{van2017neural}.

The VQ-VAE uses a codebook for  vector quantization. The codebook $\CB$ is comprised of $\log_2 |\tilde{M}|$ vectors of size $d$. The input to the quantizer $\myVec{z}$, (which is the output of the pre-quantization \ac{dnn}),  is represented by the closest codeword in $\CB$. Thus, the latent representation of  $\myVec{z}$ is 
\begin{equation}
     \hat{\myVec{z}} = \mathop{\arg\min}_{\myVec{e}_j\in\CB} \|{\myVec{z} - \myVec{e}_j}\|_{2}^{2}. 
\end{equation}

To jointly train the encoder-decoder while learning the codebook $\CB$, the  \ac{vqvae} uses a loss function comprised of three terms as follows: 
\begin{align}
\mathcal{L}_{\rm tot}(\myS ; \myX) =& \mathcal{L}(\myS ; \hat{\myS}) + \|{{\rm sg}\left(\myVec{z}\right)-\hat{\myVec{z}}}\|_{2}^{2} 
\notag \\
&
+ \beta\|{\myVec{z}-{\rm sg}\left(\hat{\myVec{z}}\right)}\|_{2}^{2},
\label{eqn:train_loss}
\end{align}
where ${\rm sg}(\cdot)$ is the stop-gradient operator.
The first term  in \eqref{eqn:train_loss},  $\mathcal{L}(\myS ; \hat{\myS})$, is the task-dependent loss (e.g., cross entropy for classification).  The second term is the VQ-loss, which   moves the codebook vectors closer to the encoder outputs. The third term is  the commitment loss, which causes the encoder outputs to be similar to the codebook vectors. The hyperparameter  $\beta >0$ balances the influence of the commitment loss on $\mathcal{L}_{\rm tot}$. 

While the loss in \eqref{eqn:train_loss} is stated for a given codebook size, it can be extended to accommodate multiple different resolutions as a form of {\em rate-adaptive} goal-oriented quantization. This is achieved by adding a regularizing term to the loss function that encourages the resulting codebook to be accommodate subsets of codewords that can be used for lower resolution quantization, see~\cite{malka2023learning}, or alternatively, by using multiple modules that quantize the quantization error of the subsequent modules, see~\cite{toderici2015variable, lei2022progressive}. 

\subsection{LLMs for Compression}
\label{subsec:llm}
Lossless data compression is a classical research area in the fields of information theory and communications. In Shannon's landmark paper ~\cite{shannon1948mathematical}, he introduced the concept of the entropy (rate) of a stationary process, and established the source coding theorem that the best performance one can achieve for data compression is the entropy rate. Shannon also discussed the English text as a particular type of data sources, and introduced the Markov process as a way to approximate it. He later posed the question on the entropy rate of English, and provided some initial estimates. This estimate was later improved by Cover and King \cite{cover1978convergent}. 

Since then, there have been considerable efforts in designing algorithms for lossless compression, leading to the development of Lempel-Ziv algorithms \cite{ziv1977universal,ziv1978compression}, the arithmetic coding algorithm \cite{rissanen1979arithmetic}, context-tree weighting algorithms \cite{willems1995context}, prediction by partial matching \cite{cleary1997unbounded,moffat1990implementing,shkarin2002ppm}, and Burrow-Wheeler transformations \cite{burrows1994block}, among others. Many of these algorithms are now embedded in computer operating systems and compression standards, such as zip-style software, JPEG standards, and MPEG standards, and become an inseparable part of everyday life. 
 
 One important part of such algorithms is a probabilistic model and the corresponding probability distribution estimation. Considering an abstract source in the alphabet $\mathcal{X}$,  the key issue here is: after we have observed past source symbols $(\ldots,x_{i-1},x_{i})$, what is the possibility distribution for the next symbol $x_{i+1}$, or the next group of symbols $(x_{i+1},\ldots,x_{i+k})$? Mathematically, we are interested in modeling the probability
 \begin{align}
 P(x_{i+1},\ldots,x_{i+k}|\ldots,x_{i-1},x_{i}).\label{eqn:pmodel}
 \end{align}
Once this is known, there are procedures to compress it to match the entropy rate, e.g., via the arithmetic coding algorithm \cite{rissanen1979arithmetic} or the enumerative coding approach \cite{cover1973enumerative}. Although many of the algorithms mentioned above do not explicitly model and learn this distribution,  the core issue in data compression is essentially how to efficiently find and utilize this distribution. 

LLMs, or more precisely, the transformer architecture \cite{vaswani2017attention} behind these models, are in fact performing the exact same task in this regard.  For a decoder-only transformer, the output logit essentially specifies the probability distribution of the next symbol $x_{i+1}$, given all the source observations in the context window of the transformer, which corresponds to the case in (\ref{eqn:pmodel}) with $k=1$. Given the wide-spread success of LLMs 
, it is reasonable to expect that they can in fact provide extremely accurate estimate for (\ref{eqn:pmodel}), and thus can lead to improved data compression performance. Indeed, in two recent works \cite{valmeekam2023llmzip,deletang2023language}, the authors showed that LLMs (LLaMA-7B and Chinchilla 70B, respectively), together with arithmetic coding, can provide superior compression performance. As noted in \cite{valmeekam2023llmzip}, the compression rate provides a significantly improved estimate of the entropy of English text, i.e., the question posed by Shannon. The authors in \cite{deletang2023language} further applied the same compression approach on natural images, which also resulted in surprisingly competitive performance, comparable to certain existing standards specifically designed for images. The authors explicitly titled their paper ``Language Modeling Is Compression'', highlighting the connection between language models and data compression.

It should be noted that there is currently diverging views on whether the parameters of the LLMs themselves should be viewed as a part of the compressed bitstream, since different from classic compression algorithms mentioned above which have no trainable parameters, LLMs have extremely large numbers of parameters. Using Shannon' formulation, the answer is clear: as long as the compressor and decompressor agree upon the parameters and do not change in the middle of compression and decompression, then the algorithms can be arbitrarily complex, and this complexity should not be counted toward the compression rate. 

This research area is still in its infancy, with many open questions to be answered. The most important question is what is the source of these significant improvements.  A few initial efforts have been made toward answering these questions. Classical studies in information theory tend to use Markov processes as a good approximation for English text. Several recent works in fact studied the transformers' ability to learn Markov models. The authors of \cite{makkuva2024attention} studied the loss landscape during transformer training on sequences generated from a single Markov chain, on a single-layer transformer. They found that there are global optimal solutions for a transformer regardless of the given Markov chain, but there are also bad local solutions for some Markov chains. Edelman et al. \cite{edelman2024evolution} studied how transformers obtain the in-context learning capability during training, and identified phase transition behaviors for transformers to form inductive heads, resulting in good in-context learning capability. In the context of data compression, the in-context learning capability essentially translates to the capability to learn the probability model (\ref{eqn:pmodel}) efficiently. We expect research connecting LLMs and compression to help further advance compression algorithms in the near future.

\subsection{Deep-Learning-Based Image/Video Transmission}
\label{sec:video}

Image and video communications are among the most popular applications of Shannon's information theory. Similar to common communication systems, an image/video communication system typically consists of three modules, each dedicated to a different purpose: source coding, channel coding, modulation. In the source coding module, inherent redundancies in image/video data are removed to minimize the amount of transmitted bits. In the channel coding module, the compressed bitstream is expanded with error-correcting codes to ensure resilient transmission against errors. The modulation module maps each small block of bits onto a   constellation point, which is associated with a complex-valued channel input symbol. 


\begin{figure}
	\centering
	\includefig{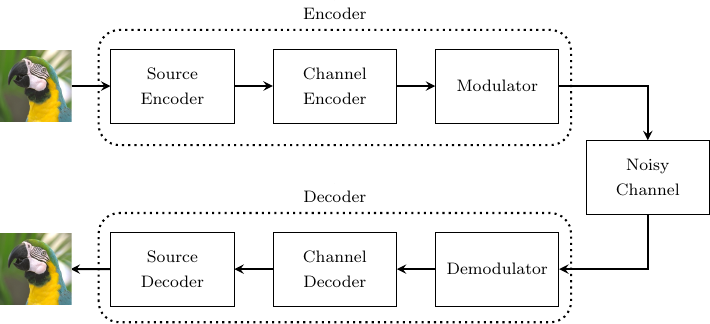}
	\caption{Separate design of source coding, channel coding, and modulation for an image/video communications system.} 
	\label{fig:jscc}
\end{figure}
\begin{figure}
	\centering
	\includefig{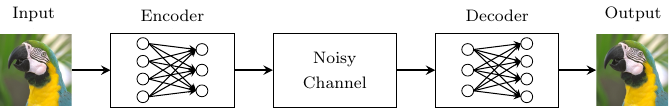}
	\caption{Joint design of source coding, channel coding, and modulation with a \ac{dnn} for an image/video communications system.} 
	\label{fig:deepjscc}
\end{figure}
Shannon proved that as blocklength approaches infinity, source coding and channel coding can be designed separately without incurring any performance loss. Therefore, a typical image/video communication system employs a two-step encoding process, as illustrated in Fig.~\ref{fig:jscc}. However, it is well known that separate source-channel coding (SSCC) is not optimal for finite blocklengths. In such cases, JSCC often achieves better performance in practice.  In \cite{8723589}, a pioneering deep-learning-based JSCC system for wireless image transmission, named DeepJSCC, was proposed. DeepJSCC can be viewed as an auto-encoder comprising two jointly trained convolutional neural networks (CNNs)—one functioning as the encoder and the other as the decoder 
(see Fig.~\ref{fig:deepjscc}).  It directly maps image pixels onto constellation points, bypassing traditional source and channel codes used for data compression and error correction. Experimental results demonstrate that DeepJSCC significantly outperforms traditional image transmission systems that concatenate a JPEG or JPEG2000 encoder with a capacity-achieving channel code, especially at low signal-to-noise ratio (SNR) and channel bandwidth values for additive white Gaussian noise channels. More importantly, DeepJSCC does not suffer from the {\em cliff effect}, as it provides a graceful degradation in performance when the real-world channel SNR deviates from the assumed training SNR. 
For slow Rayleigh fading channels, DeepJSCC also significantly outperforms traditional separate image communication systems across all SNR and channel bandwidth values.

It is well known that feedback does not increase the capacity of memoryless channels, but can simplify coding and improve reliability in the  finite blocklength regime.  Encouraged by the success of DeepJSCC, \cite{9066966} investigates how noiseless or noisy channel feedback can enhance the reconstruction quality at the receiver in an image communication system. This system, an upgraded version of DeepJSCC, is termed DeepJSCC-feedback (DeepJSCC-f). Utilizing modern machine learning techniques, DeepJSCC-f significantly improves end-to-end reconstruction quality for fixed-length transmissions and reduces average delay for variable-length transmissions. DeepJSCC-f is the first practical JSCC system that fully exploits channel feedback to outperform traditional communication systems.

Layered coding and multiple description coding are two important problems in information theory. Both involve compressing a source into multiple bitstreams that can be combined by the receiver to enhance the quality of the reconstructed source. However, these two coding schemes differ significantly. In layered coding, the bitstreams are sent sequentially, and each layer incrementally refines the reconstructed source. In contrast, multiple description coding involves bitstreams that are independent of each other, allowing them to be retrieved in any order. In \cite{9464731}, building on the foundation of DeepJSCC, the DeepJSCC-layer (DeepJSCC-l) was proposed as a deep learning-based communication system for adaptive-bandwidth transmission of images over wireless channels. DeepJSCC-l features three architectures tailored to different complexity tradeoffs. Remarkably, compared to single-layer or single-description transmission, the end-to-end performance loss of DeepJSCC-l is almost negligible. Additionally, DeepJSCC-l inherits the desired property of graceful degradation with varying channel SNR.

Though DeepJSCC and its variants, DeepJSCC-f and DeepJSCC-l, yield impressive results for wireless image transmission, they focus solely on the objective distortion of reconstructed images relative to the input image, neglecting their subjective perception by humans. To address this, \cite{10158995} introduced DeepJSCC-perception (DeepJSCC-p), which leverages the perceptual quality enhancements of deep generative models for wireless image transmission. DeepJSCC-p includes two versions: InverseJSCC and GenerativeJSCC. InverseJSCC denoises the outputs of DeepJSCC by solving an inverse optimization problem using a pre-trained style-based generative adversarial network (StyleGAN). GenerativeJSCC, on the other hand, involves end-to-end training of an encoder and a StyleGAN-based decoder. Both versions measure objective distortion with \ac{mse} and subjective perceptual quality with learned perceptual image patch similarity. Compared to DeepJSCC, DeepJSCC-p significantly enhances the perceptual quality of reconstructed images.



Parallel to the advancements in DeepJSCC for image transmission, \cite{9837870} introduced DeepWiVe, the first end-to-end JSCC system for video transmission. DeepWiVe utilizes DNNs to directly map video pixels onto  constellation points, integrating video compression, channel coding, and modulation  into a single DNN.





\section{Summary and Road Ahead}
\label{sec:concl}
Recent theoretical advances have led to the development of concrete algorithmic tools, establishing the foundations for the framework of  {\em goal-oriented quantization}. As discussed in Sections \ref{ssec:GOQaunt} and \ref{subsec:Neural-Goal}, goal-oriented quantization systems can be designed using model-based approaches, which rely on mathematical representations of the underlying task, or data-driven approaches, where the quantization mapping is learned from data. The inherent limitation of the former to simplified mathematical models (e.g., linear and quadratic tasks) makes  data-driven approaches, particularly those aided by \ac{dnn}, highly attractive.  
Nonetheless, resorting to purely data-driven designs gives rise to challenges associated with black-box solutions, in terms of interpretability, complexity, and generalization. 
A promising approach to leverage the advantages of both model-based and data-driven designs is through the emerging paradigm of model-based deep learning~\cite{shlezinger2022model,shlezinger2020model}. Future research could explore the use of \ac{dnn}s not to replace interpretable goal-oriented quantization designs derived from mathematical task representations but to integrate data-driven tools to address limitations in more complex tasks.

Rate-distortion-perception theory reviewed in Section \ref{subsec:rdp} generalizes  Shannon's original framework  by incorporating a perception constraint alongside the distortion constraint.
This theory is rapidly evolving both in terms of its formulation and content. 
Note that the role of a perception constraint is to ensure that the reconstruction distribution does not significantly deviate from the source distribution. However, in practice, there is often a need for more fine-grained control over the reconstruction distribution. Additionally, in many cross-domain compression problems, such as those arising from image restoration tasks, the objective goes beyond enforcing distributional consistencies. This motivates the study of
output-constrained lossy source coding \cite{saldi2015output,liu2022lossy}, where the source sequence  is transported 
through an information bottleneck to produce a new sequence  with a prescribed distribution at the minimum end-to-end distortion. This line of research broadens the scope of perception-aware lossy source coding and has the potential to establish a new paradigm for image compression and restoration by integrating principles from information theory and transport theory.

Compression for distributed estimation and learning is a vast subject with numerous exciting developments in both theory and practice. In Section \ref{subsec:est}, we have only been able to provide a very partial coverage of this expansive field.
Future directions  include developing scalable and efficient algorithms,  advancing privacy-preserving techniques,  designing adaptive compression methods, enhancing robustness to adversarial attacks, and strengthening theoretical foundations.

New neural network models, such as transformers, can produce accurate probability distribution estimates. 
As discussed in 
Section \ref{subsec:llm}, one notable application of this capability is the use of LLMs for lossless data compression. This emerging research direction presents many open questions, including understanding the exact mechanisms behind the observed improvements, developing methods to reduce algorithm complexity, and exploring additional functionalities such as fault tolerance.

Traditionally,  image/video communications systems employ separate designs for the image/video codec and the channel codec. The advent of deep learning has inspired researchers to consolidate these components, along with the modulator and demodulator, using trainable neural networks. As shown in Section \ref{sec:video}, this joint design approach significantly outperforms the traditional separate design and can be further enhanced for channels with feedback. This new paradigm offers a promising research direction, with potential extensions to various problems in information theory, such as layered coding, multiple description coding, and distributed source coding~\cite{dsc1,dsc2,dsc3}.

\bibliographystyle{IEEEtran}
\bibliography{IEEEabrv,refs,tian,fang}

\end{document}